# Photoelectron backscattering in vacuum phototubes

B.K.Lubsandorzhiev[*], R.V.Vasiliev, Y.E.Vyatchin, B.A.J.Shaibonov.

*Institute for Nuclear Research of RAS*

[*]Corresponding author:
Tel: +7(095)1353161; fax: +7(095)1352268;
e-mail: lubsand@pcbai10.inr.ruhep.ru
Address: 117312 Moscow Russia, pr-t 60-letiya Oktyabrya 7A
Institute for Nuclear Research of RAS

**Abstract**

In this article we describe results of studies of a photoelectron backscattering effect in vacuum phototubes: classical photomultipliers (PMT) and hybrid phototubes (PH). Late pulses occurring in PMTs are attributed to the photoelectron backscattering and distinguished from pulses due to an anode glow effect. The late pulses are measured in a number of PMTs and HPs with various photocathode sizes covering 1-50 cm range and different types of the first dynode materials and construction designs. It is shown that the late pulses are a generic feature of all vacuum photodetectors – PMTs and PHs and they don't deteriorate dramatically amplitude and timing responses of vacuum phototubes.

## 1. Introduction

Vacuum phototubes are used widely in an overwhelming majority of experiments in astroparticle and high energy physics. Precision timing of vacuum phototubes plays a crucial role in defining detectors angular and amplitude resolutions, background suppression etc. Phototube timing performance is of particular importance in case of experiments dealing with extremely low intensities of light fluxes, large-scale Cherenkov experiments in particular. One of the most strongly influencing effects on phototube timing are so called "late" pulses and prepulses [1-4]. The late pulses are attributed to the photoelectron backscattering effect on the first dynodes of classical PMTs or the anode structures of HPDs [2,3]. This effect smears not only the timing response of vacuum phototubes but their amplitude resolution too [5].

## 2. Late pulses

It is still important to make distinction between the late pulses and afterpulses. Afterpulses stem from ionisation of the residual gas atoms and the atoms adsorbed by the first dynode surface, or luminescence of the dynodes and the residual gas [6-8]. Afterpulses are always correlated with the main pulses. The delay time of the ion-feedback afterpulses for classical PMTs stretches from hundred nanoseconds to dozens of microseconds. In our study the afterpulses nearest to the main pulses are suppressed completely by the discriminator deadtime (~200ns).

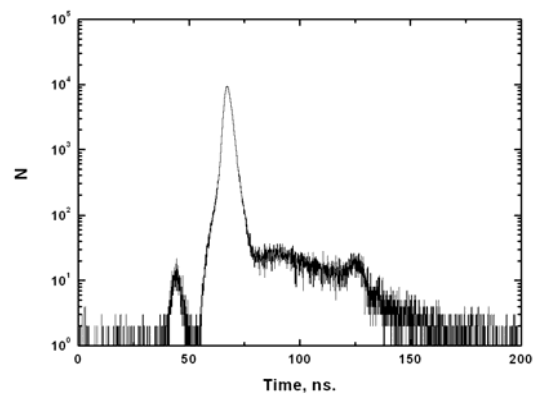

Fig.1. Single photoelectron transit time distribution of EMI9350

The late pulses are in fact a part of the main pulses of the phototube's response but they are only delayed by less than ten nanoseconds in small phototubes and several dozens of nanoseconds in large phototubes. As it was mentioned above it is supposed that the late pulses arise from photoelectron backscattering (elastically or inelastically) on the first dynodes of classical PMTs or the anode structures of HPDs (silicon diodes or luminescent screens). A photoelectron hitting the first dynode may be backscattered even without liberating any secondary electrons. In turn backscattered photoelectrons are decelerated by the electric field and then accelerated again towards the electron multiplying system producing finally the phototube's output signal. Thus the resulting delay time may be up to twice the photoelectron transit time between the photocathode and the electron multiplying system (the first dynode in case of PMT). We introduced in our previous work [3] the late pulses probability coefficient $K_{late}=K_{el}+K_{in}$, where $K_{el}$ and $K_{in}$ are the elastic and inelastic backscattered late pulses probability coefficients respectively.

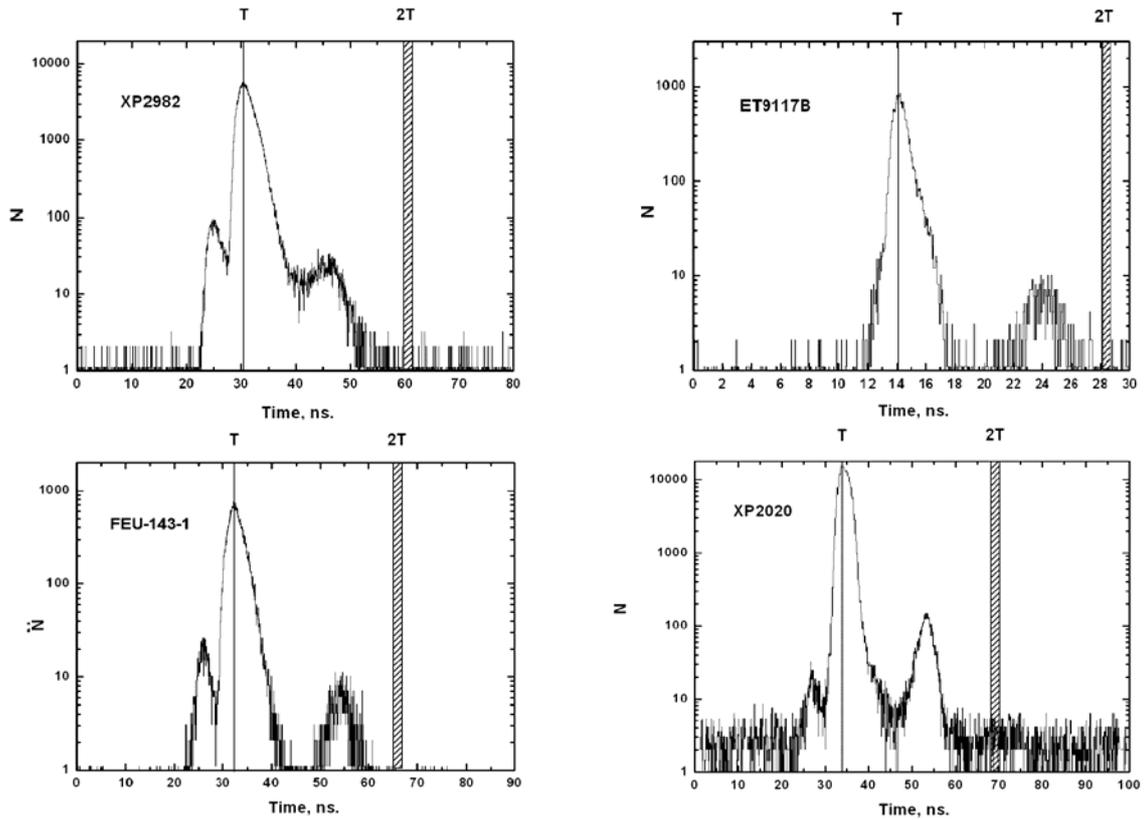

Fig.2. Single photoelectron transit time distributions of a number of classical PMTs: clockwise from the upper left XP2982, ET9117B, FEU-143-1, XP2020.

The typical photoelectron transit time distribution of the 8" EMI9350 phototube is shown in Fig.1. The distribution was measured with the discriminator threshold of 0.01 p.e.. The first peak of the distribution is due to prepulses [3]. The second "main" peak (between 50 and 70 ns) corresponds to the main pulses. The time interval between the prepulses peak and the main peak corresponds to the photoelectron transit time from the photocathode to the first dynode $t_{c-1d}$. The part of the distribution with time entries of more than 75 ns is explained largely by the late pulses. The broad peak around 110 ns with tail up to 140 ns is very likely due to the inelastically backscattered photoelectrons. The third rather sharp peak at 150 ns is attributed to the elastically backscattered photoelectrons. The time interval between the second and third peaks amounts roughly to twice the value of $t_{c-1d}$.

We measured the photoelectron transit time distributions for a number of various types of PMTs and hybrid phototubes which differ very much in their sizes and designs. Some typical distributions for classical PMTs are shown in Fig.2. All distributions were measured with 0.1 p.e. threshold and in all of them the late pulses peaks are clearly seen. The vertical lines and bars mark the mean value of the overall electron transit times (**T**) of PMTs and the doubled mean transit times (**2T**) respectively. One can see in Fig.2 that the third peaks in all distributions stay well apart from **2T**.

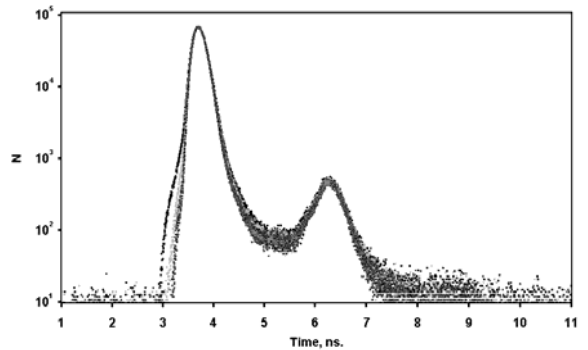

Fig.3. Single photoelectron transit time distribution of H7422 [9]. Courtesy of Becker&Hickl.

Fig.3 and Fig.4 show the single photoelectron transit time distributions for HAMAMTSU phototubes H7422 and R3809U respectively. The former is a PMT module with metal channel dynodes and the latter is a PMT-MCP. The distributions are borrowed from [9]. As in the case of classical PMTs, the peaks of the late pulses are distinctly seen. The late pulses contribution to the total distributions is practically the same as for classical PMTs. In Fig.3 four distributions measured with four thresholds of the discriminator are shown.

The single photoelectron transit time distribution of the Quasar-370 phototube [10-12] is shown in Fig.5. the Quasar-370 is a hybrid phototube using luminescent screen as the first stage of the photoelectron multiplication. There is no late pulses peak in the single photoelectron transit time distribution but backscattered photoelectrons are seen as rather sharp peaks in the phototube's time response, Fig.6. The distances between peaks are of ~20 ns - twice the photoelectron transit time from the photocathode to the luminescent screen. The phototube was illuminated by short, ~ 0.5 ns width, light pulses from a nitrogen laser.

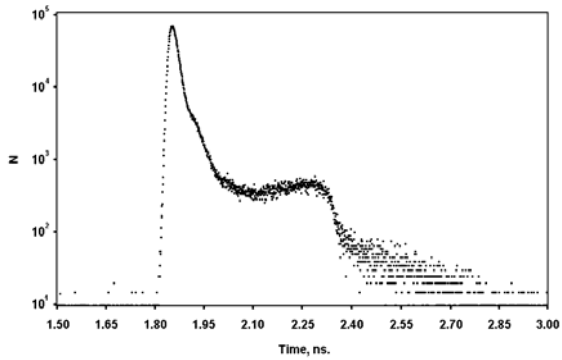

Fig.4. Single photoelectron transit time distribution of PMT-MCP R3809U [9]. Courtesy of Becker&Hickl

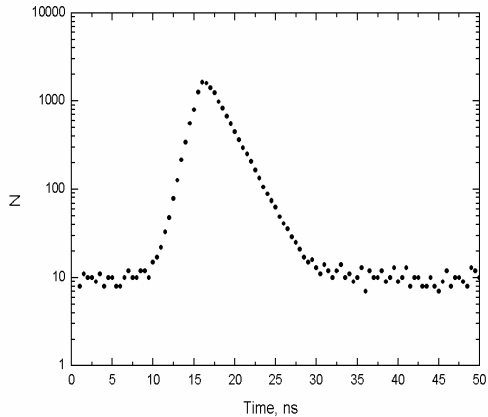

Fig.5. Single photoelectron transit time distribution of Quasar-370.

The data of our study are listed in table 1. One can see from the data that, regardless of the phototube types, first dynode materials etc, the values of the late pulses probability coefficient $K_{late}$ are rather small and do not exceed 2%. On the other hand, the backscattering probability coefficient $\eta$ depends strongly on $Z_{eff}$ and, for $Z_{eff} = 54$, $\eta$ is larger than 40% [13, 14]. The questions arise. Why is $K_{late}$ so small in comparison with $\eta$? Why $K_{late}$ does not increase with increasing $Z_{eff}$? For the Quasar-370 phototube, the situation is very intriguing: the value of $Z_{eff}$ is rather high (39) but $K_{late} = 0$! It is interesting that there is even no peak due to the photoelectron elastic backscattering. So where did backscattered photoelectrons disappear?

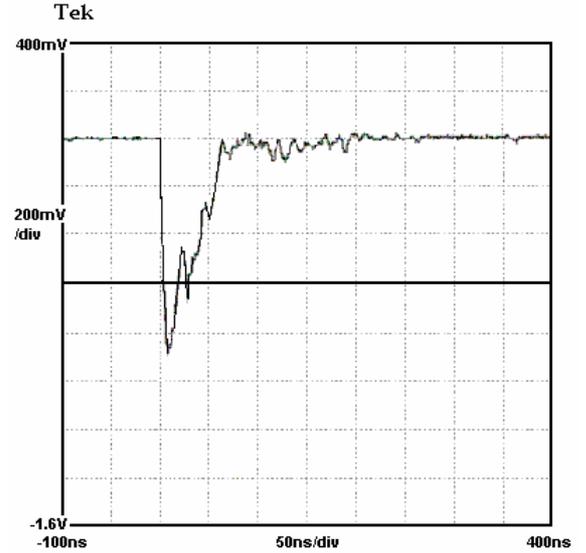

Fig.6. Waveform of output signal of Quasar-370

One can reconcile experimental data on $K_{late}$ and $\eta$ by recalling the secondary electron emission (SEM) phenomenological models and experimental data on SEM. The overwhelming bulk of so called "genuine" secondary electrons is produced by the backscattered photoelectrons rather than the direct ones. So the more the backscattering probability coefficient $\eta$ the more the secondary emission coefficient $\sigma$ [13,14]. The absence of the late events peak in the single photoelectron transit time distribution for the Quasar-370 phototube is explained by the fact that the inelastically backscattered photoelectrons produce enough photons in the luminescent screen to be registered by the small PMT of the phototube. The elastic backscattering probability depends on the initial energy of electrons as $K_{el} \sim E^{-A}$, where **A** is a constant,: at **E = 25 keV** the value of $K_{el}$ is less than 0.1% [13].

The single photoelectron charge distribution of the R1463 tiny classical PMT is presented in Fig.7. It is, in a sense, a quite unique PMT. One can see not only the very nice single photoelectron peak around channel #140 but also peaks generated by prepulses on the first and even second dynodes, the second and first peaks on the left side of the spectrum respectively. Switching off the photocathode it is possible to measure prepulses spectrum directly, see grey curve in Fig.7. Subtracting the prepulses spectrum from the total one we get a clear single photoelectron charge spectrum without the exponentially rising left part which has been plaguing experimentalists for many years. It seems in this case that the late pulses result in very small distortion of the left part of the spectrum.

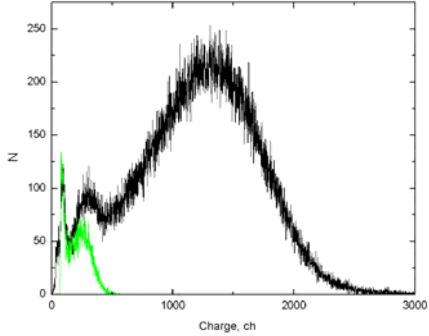

Fig.7. Charge distributions of single photoelectron pulses (black curve) and prepulses (grey curve) of R1463.

It is worth noting that there is a non-negligible probability $K_{miss}$ for the backscattered photoelectrons to miss hitting the first dynode a second time depending on the transverse momentum and the electric field configuration in the front–end geometry of the PMTs. As it might be inferred from the results reported here, it seems plausible that the value of $K_{miss}$ is quite small. It is rather difficult to estimate $K_{miss}$ experimentally but it can be done by some computer simulations. To know exactly the value of $K_{miss}$ is of particular importance because it is closely related with the determination of the effective quantum efficiency in vacuum phototubes.

### III. Anode glow

It has long been known that electrons in classical PMTs induce light emission in the last cascades of dynode systems – the "anode glow" [15]. In some cases, photons of the glow could generate peaks in the photoelectron transit time distributions very similar to the late pulses peaks but the time interval between the main peak and the peak due to the anode glow should be equal to the overall electron transit time of PMT [16].

To distinguish the photoelectron backscattering peak from the anode glow peak we measured the anode glow directly by using the small PMT (XP2020) viewing right at the anode region of the PMT under study (EMI9350) from outside. It is known that the most part of the emission spectrum of the anode glow lies in the spectral region from 400 nm to 700 nm so that the PMT's glass bulb is transparent to the anode glow photons [17]. In Fig.8 the single photoelectron transit time distribution of the PMT under study and the anode glow kinetics recorded by the small PMT are shown. Here all existing time delays are taken into account. It is clearly seen that peaks due to the photoelectron backscattering and the anode glow don't coincide with each other. The time difference between them is about 12 ns. Moreover it should be reminded here that pulses due to the anode afterglow are most likely suppressed completely by the dead time of the discriminator used in the measurements. Indeed, the discriminator threshold in the measurements was rather low (~0.01) p.e. so the photoelectrons initiating the anode glow were very likely registered and the electronic system was insensitive to delayed pulses in the time domain of 200 ns. Moreover from Fig.2 it is clear that the late events arrive conspicuously earlier than the doubled overall transit time of PMTs. So it seems one can say with confidence that the late events peak is not due to the anode glow.

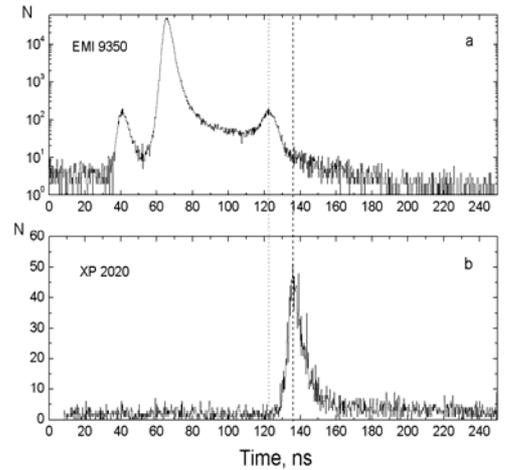

Fig.8. Single photoelectron transit time distribution of ET9350 (a) and the PMT's anode glow kinetics (b).

### IV. Conclusion

The problems of precision timing and effective quantum efficiency of vacuum phototubes are deeply interconnected. One aspect common to both problems is the phototelectron backscattering effect. The photoelectron backscattering is the generic feature inherent to all types of vacuum photodetectors. There are no substantial discrepancies between experimental data on the electron backscattering probability $\eta$ and the late pulses probability coefficient $K_{late}$. It seems that photoelectron backscattering does not deteriorate dramatically the amplitude and timing resolution of vacuum photodetectors. Further, elaborate joint experimental and computer simulation efforts are necessary to fully understand the photoelectron backscattering effect and to shed light on the vacuum phototubes effective quantum efficiency problem.

### Acknowledgements

Authors would like to thank Drs. W.Becker and A.Bergmann for kindly allowing to cite their results and Dr.V.Ch.Lubsandorzhieva for careful reading the paper and many valuable remarks.

Table 1. Late pulses and prepulses probabilities coefficients for a number of vacuum phototubes

| PMT | 1st dynode material | Cathode diameter mm | $\sigma$ | $Z_{eff}$ | $K_{pre}$, % | $K_{late}$, % |
|---|---|---|---|---|---|---|
| H7422 | Metal-channel | 5 | 4-6 | * | 1 | 1 |
| R1463 | NaKCsSb | 10 | 4-6 | 50 | 4 | 0.7 |
| R3809 | NaKCsSb | 11 | 3 | 79 | <0.1 | 1 |
| EMI9083B | $SbCs_3$ | 15 | 6-10 | 54 | 1.1 | 0.4 |
| EMI9116B | $SbCs_3$ | 22 | 6-10 | 54 | 1 | 1 |
| XP2982 | CuBe | 23 | 5 | 28 | 1 | 1 |
| FEU-130 | GaP | 25 | 20 | 29 | 0.3 | 1 |
| FEU-"Baikal-1" | AlMg | 25 | 3 | 12.5 | <0.1 | 0.6 |
| EMI9117B | $SbCs_3$ | 32 | 6-10 | 54 | 1 | 1 |
| FEU-143-1 | GaP | 40 | 20 | 29 | 2.3 | 1 |
| XP2020 | CuBe | 44 | 5 | 28 | 1 | 1 |
| EMI9350 | $SbCs_3$ | 190 | 6-10 | 54 | 1 | 1.5 |
| R1449 | CuBe | 460 | 3-5 | 28 | 1 | 1 |
| Quasar-370 | $Y_2SiO_5$:Ce | 370 | 25 | 39 | 0 | 0 |

$\sigma$ - secondary emission coefficient of first dynode; $Z_{eff}$ –effective atomic number of first dynode material; $K_{pre}$ – prepulses probability coefficient; $K_{late}$ – late pulses probability coefficient; * - no data.

**Figures:**

Fig.1. Single photoelectron transit time distribution of EMI9350

Fig.2. Single phototelectron transit time distributions of a number of classical PMTs: clockwise from the upper left XP2982, ET9117B, FEU-143-1, XP2020.

Fig.3. . Single photoelectron transit time distribution of H7422 [9]. Courtesy of Becker&Hickl

Fig.4. . Single photoelectron transit time distribution of PMT-MCP R3809U [9]. Courtesy of Becker&Hickl

Fig.5. Single photoelectron transit time distribution of Quasar-370.

Fig.6. Waveform of output signal of Quasar-370

Fig.7. Charge distributions of single photoelectron pulses (black curve) and prepulses (grey curve) of R1463

Fig.8. Single photoelectron transit time distribution of EMI9350 (a) and the PMT's anode glow kinetics (b)